\documentclass[fleqn,usenatbib]{mnras}

\usepackage{newtxtext,newtxmath}

\usepackage[T1]{fontenc}

\DeclareRobustCommand{\VAN}[3]{#2}
\let\VANthebibliography\thebibliography
\def\thebibliography{\DeclareRobustCommand{\VAN}[3]{##3}\VANthebibliography}

\usepackage{graphicx}	
\usepackage{amsmath}	
\usepackage{tabularx}
\usepackage{afterpage}
\usepackage{pdflscape}   
\usepackage{booktabs}    
\usepackage{caption}
\usepackage{subcaption}
\usepackage{lineno}
\usepackage{rotating}

\newcommand{\Msun}{\mbox{$\mathrm{M}_{\odot}$}}

\newcommand{\edit}[1]{\textcolor{black}{#1}}
\newcommand{\editmore}[1]{\textcolor{black}{#1}}

\title[Planetary hydrogen and WD spectral evolution]{Planetary material is a major source of hydrogen in helium-atmosphere white dwarfs}

\author[P.~Izquierdo, A.~Swan et al.]{Paula~Izquierdo,$^{1}$\thanks{Joint first authors}\thanks{Email: paula.izquierdo@warwick.ac.uk}
Andrew~Swan,$^{1}$\thanks{Email: andrew.swan.astro@gmail.com}\footnotemark[1]
Christopher~J.~Manser,$^{2}$
Boris~T.~G\"ansicke,$^{1}$
Joan~Najita,$^{3}$
\newauthor
Jessica N. Aguilar,$^{4}$
Steven Ahlen,$^{5}$
Leandro {Beraldo e Silva,}$^{6,7}$
Davide Bianchi,$^{8,9}$
David Brooks,$^{10}$
\newauthor
Todd Claybaugh,$^{4}$
Kyle Dawson,$^{11}$
Axel de la Macorra,$^{12}$
Peter Doel,$^{10}$
Andreu Font-Ribera,$^{13,14}$
\newauthor
Jaime E. Forero-Romero,$^{15,16}$
Enrique Gazta\~{n}aga,$^{17,18,19}$
Satya Gontcho A Gontcho,$^{20}$
Gaston Gutierrez,$^{21}$
\newauthor
Klaus Honscheid,$^{22,23,24}$
Robert Kehoe,$^{25}$
Sergey E. Koposov,$^{26,27}$
Andrew Lambert,$^{4}$
Martin Landriau,$^{4}$
\newauthor
Laurent Le Guillou,$^{28}$
Ting S. Li,$^{29}$
Aaron Meisner,$^{3}$
Ramon Miquel,$^{13,14}$
John Moustakas,$^{30}$
\newauthor
Francisco Prada,$^{31}$
Ignasi P\'erez-R\`afols,$^{32}$
Alexander H. Riley,$^{33}$
Graziano Rossi,$^{34}$
Eusebio Sanchez,$^{35}$
\newauthor
Michael Schubnell,$^{36,37}$
David Sprayberry,$^{3}$
Gregory Tarl\'{e},$^{37}$
Benjamin A. Weaver,$^{3}$
Hu Zou,$^{38}$
\\
Affiliations are listed at the end
}

\date{Accepted XXX. Received YYY; in original form ZZZ}

\pubyear{\the\year{}}

\begin{document}
\label{firstpage}
\pagerange{\pageref{firstpage}--\pageref{lastpage}}
\maketitle

\begin{abstract}
Many white dwarfs with helium-dominated atmospheres show traces of hydrogen in their spectra. On average, the total mass of hydrogen within their convective envelopes increases with age, but there is no consensus on the origin of such hydrogen. Hypotheses include upward diffusion from hidden reservoirs deep within the envelope, or accretion, alongside heavier elements, of planetary material. We construct a simple model that includes both pathways, and test it using two independent samples of helium-atmosphere \editmore{DB} white dwarfs from the DESI and SDSS spectroscopic surveys\edit{, selected to have \textit{Gaia} colours $G_{\rm{BP}}-G_{\rm{RP}}>- 0.25$, corresponding roughly to 10\,$000\lesssim T_{\textrm{eff}}\lesssim19$\,000\,K}. We find that at least $22\pm2$ per cent of the helium-atmosphere white dwarfs with detectable hydrogen must have accreted some of it from their planetary systems, via ices, hydrated minerals or atmospheres. We thus demonstrate that planetary hydrogen is a necessary ingredient to explain the spectral evolution of white dwarfs as a population, rather than only extreme hydrogen-enriched cases.
\end{abstract}




\section{Introduction}
\label{sec:introduction}

White dwarfs are dense stellar remnants that have ceased nuclear fusion and whose destiny is to cool and crystallise \edit{\citep{Mestel1952, Salpeter1961, Tremblay2019}}. Their intense gravitational fields \citep{Schatzman1958} force atoms to stratify by mass, producing in most cases a core of carbon and oxygen surrounded by a helium-rich envelope and a thin outer layer of hydrogen \citep{Althaus2010}. The interplay between processes such as diffusion \citep{Unglaub1998}, stellar winds \citep{Unglaub2000} and convection \citep{MacDonald1991,Rolland2018} determines the atmospheric composition of the star, and thus its spectral type. If the balance between these processes shifts, a white dwarf can change its appearance \citep{Bedard2024}. Similarly, if it begins accreting material from its environment, changes to its atmospheric composition may be detectable spectroscopically.

Most newly formed white dwarfs have sufficient amounts of \textit{primordial} hydrogen (i.e. inherited from their prior evolution) to have hydrogen-dominated atmospheres during their whole lives, thus preserving their original spectral classification at all cooling ages. However, the fraction of helium-dominated atmosphere white dwarfs varies along the cooling sequence: they account for around 24 per cent of white dwarfs with cooling age $\sim1$\,Myr, falling to only about 8 per cent at $\sim10$\,Myr, rising again as high as 40 per cent when they reach several Gyr \edit{\citep[e.g.][]{Bedard2020,Cunningham2020,LopezSanjuan2022,JiminezEsteban2023,Torres2023,OBrien2024_40pc,Kilic2025, Kilic2026cool}}. This phenomenon is termed spectral evolution and is linked to changes in the atmospheric hydrogen abundance.

He-atm~WDs\footnote{For readability, we introduce the shorthand ``He-atm~WD'' to mean ``white dwarf with a helium-dominated atmosphere''. Such stars may have detectable traces of hydrogen (``He+H-atm~WD''), metals (``He+Z-atm~WD''), or both (``He+H+Z-atm~WD''). Unless obvious from the context, references to He-atm~WDs or He+H-atm~WDs concern both the classes and their sub-types.} are formed via a late helium flash that burns almost all remaining hydrogen while a star is on the asymptotic giant branch \citep{Iben1983,Herwig1999}, or later via stellar mergers \citep{Zhang2012}. As they cool, they develop convection zones that encompass the atmosphere and envelope, and efficiently mix the material within them. Some He-atm~WDs show traces of hydrogen (He+H-atm~WDs) during their whole lifetime, while other He+H-atm~WDs evolve from previously hydrogen-dominated white dwarfs: convection zones get bigger as stars age, diluting thin superficial layers of hydrogen \edit{\citep{Fontaine1987, Koester2015, Rolland2018}}, causing such hydrogen-dominated stars to transition into He+H-atm~WDs.

Independently of the formation channel, total hydrogen masses in He+H-atm~WD convection zones monotonically increase with cooling age  \citep{Bergeron2011, Raddi2015}. A theoretical ``float-up'' scenario has been proposed to explain this hydrogen increase \citep{Rolland2020}, in which the deep, non-convective helium envelope may contain a reservoir of primordial hydrogen that diffuses upwards on long timescales, and is mixed in to the atmosphere as the convection zone extends downwards. Models that include this diffusion can reproduce atmospheric hydrogen abundances observed in most of the He+H-atm~WDs \citep{Bedard2023}. 

However, there is a positive correlation between the detection of hydrogen and metals in the atmospheres of He-atm~WDs \citep{Kilkenny1986,Gianninas2004,Farihi2013,Kawka2005,GentileFusillo2017,Hoskin2020,Izquierdo2021GD424,Izquierdo2026DxZ}. Also some He+H-atm~WDs have extremely high hydrogen abundances \citep{Gianninas2004,Koester2005DAZB}. Neither of these observations can be explained by the float-up scenario \citep{Rolland2020}. Thus, an additional, time-dependent source of hydrogen appears to be involved in the spectral evolution of \edit{some} He+H-atm~WDs. 

Historically, the interstellar medium was considered as a potential source for external metal-enrichment of white dwarfs \citep{Alcock1980a}, before accretion from remnant planetary systems was proposed as an alternative \citep{Alcock1986}. The interstellar scenario has since been decisively ruled out in favour of planetary systems, which can supply both hydrogen and metals \edit{\citep{Farihi2010rockyPlanetesimals, Zuckerman2010}}. About 25--50 per cent of all white dwarfs are observed to accrete metals \citep{Zuckerman2003,Koester2014,Manser2024HePlanets,OuldRouis2024} but, in most cases, that material appears dry and rocky \citep{Trierweiler2023}, adding only traces of hydrogen. However, \edit{water can survive within planetesimals throughout post-main-sequence evolution \citep{Jura2010,Malamud2016waterI,Malamud2017waterII,Malamud2017waterIII}, and} the number of systems thought to be accreting large quantities of hydrogen is growing, with sources including water-rich asteroids  \citep{Farihi2013,Raddi2015}, icy objects  \citep{Xu2017,Klein2021,Johnson2023,Sahu2025ice} and giant planets \citep{Gansicke2019,Schreiber2019}. Crucially, while metals sink out of the convective envelope, accreted hydrogen accumulates.

\edit{It has been proposed that a large fraction of stars with trace hydrogen may have accreted water-rich debris at some point in their history \citep{Raddi2015}, as some heavily metal-enriched He-atm stars have large quantities of hydrogen, and excess photospheric oxygen after accounting for the metal oxides in dry rock. Hydrogen is detected almost twice as frequently at He-atm white dwarfs with photospheric metals than at those without, further supporting the hypothesis that accretion of planetary hydrogen is involved \citep{GentileFusillo2017}.} In this paper we model the accretion of planetary material onto He-atm WDs and \edit{set a stringent lower limit on} the fraction of He+H-atm~WDs whose hydrogen includes a planetary contribution.

\section{Modelling the origin of hydrogen}
\label{sec:model}

\begin{figure}
\centering
\includegraphics[width=\hsize]{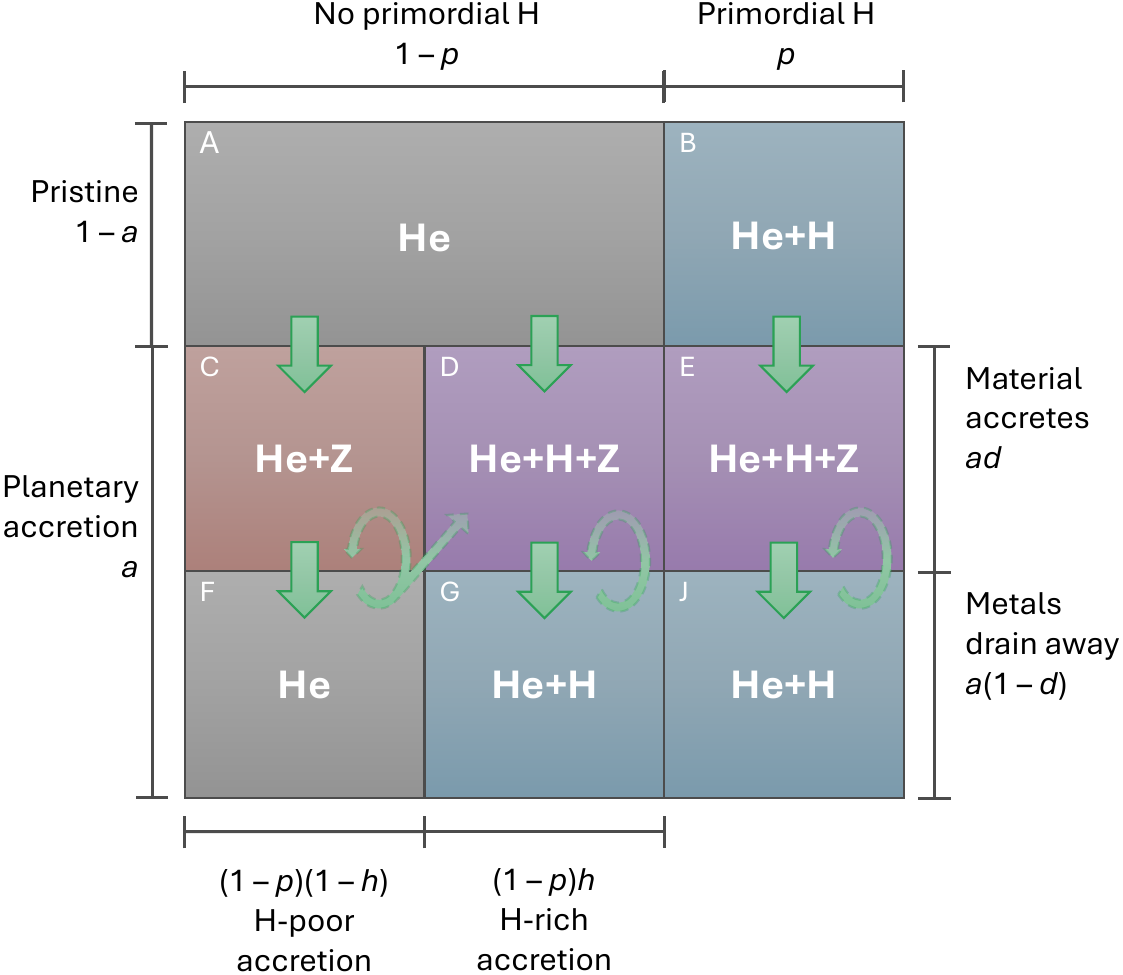}
\caption{Evolutionary phases and pathways of He-atm~WDs considered in our model. Distinct phases are represented by cells A--J, whose labels and colours indicate their spectral features. Stars in the first row have never accreted planetary material (`pristine'), and their \textit{primordial} hydrogen may or may not be detectable (cells B and A, respectively). If and when they accrete planetary material, they move downwards along the pathways shown as green arrows, and cannot return to the top row. The middle row corresponds to He-atm~WDs that have \textit{detectable} atmospheric pollution. For He-atm~WDs with no primordial hydrogen, the accreted material may be hydrogen-poor or hydrogen-rich (C and D cells, respectively). The third row contains He-atm~WDs that have accreted planetary material, but whose metals have sunk out of their photospheres and are thus untraceable. Cell side lengths are labelled with the parameters of our model, as described in the main text. While the arrows show potential evolutionary pathways, it is critical to understand that our model is applied to a population snapshot, where the number of stars in a given phase at the moment of observation is static.}
\label{fig:boxes}
\end{figure}

We construct a simple model of planetary accretion onto He-atm~WDs, as shown in Fig.~\ref{fig:boxes}. Each cell in the schematic diagram represents an evolutionary phase for He-atm~WDs, and the associated spectral features. The top row represents stars with pristine atmospheres that have not experienced accretion (A,B). They evolve downwards if accretion begins, enter the middle row (C, D, E) if atmospheric pollution reaches detectable levels, and ultimately progress to the bottom row once metals have drained away (F, G, J). 

Accretion of hydrogen-poor material (e.g. dry rock) will only affect detectability of metals, regardless of whether the star has primordial hydrogen (B\,$\rightarrow$\,E) or not (A\,$\rightarrow$\,C). 

Accretion of hydrogen-rich material by a white dwarf with undetectable primordial hydrogen can raise its hydrogen abundance above the detection threshold (A\,$\rightarrow$\,D). In stars with primordial hydrogen, accretion of such material will increase the total amount of hydrogen, but the only change in spectral appearance will be from metals (B\,$\rightarrow$\,E).

The detection of metals implies that we observe stars in an ongoing, or recent phase of accretion \citep{Koester2009}. In systems where accretion ceased a sufficiently long time ago, the metals will have diffused out of the envelope, causing stars that were born without primordial hydrogen and that did not accrete any noticeable amounts of hydrogen-rich material to revert to being pure He-atm~WDs (F). In contrast, stars that accreted hydrogen-rich material will appear as He+H-atm~WDs (G and J) independently of whether they were born with primordial hydrogen or not. Stars can cycle from F to C, G to D, or J to E once a new accretion episode begins. Stars can also migrate from cells C or F to cell D once sufficient hydrogen has been accreted to exceed the detection threshold.

The arrows in Fig.~\ref{fig:boxes} show various evolutionary pathways that He-atm~WDs may follow as they accrete planetary material, and thus how each cell comes to be populated. However, our data represent a snapshot of the whole He-atm~WD population, frozen in time, so our model has no time-dependence. We only need to consider which cell a given star currently inhabits and how accretion impacts the population \textit{in its current state}.

Given that the total area of cells A--J in Fig.~\ref{fig:boxes} accounts for the whole population of He-atm~WDs, we can use the fractional area of each cell to derive the fraction $f$ of He+H-atm~WDs whose hydrogen must include a planetary contribution:

\begin{equation}
f=\frac{\textrm{D}+\textrm{G}}{\textrm{B}+\textrm{D}+\textrm{E}+\textrm{G}+\textrm{J}}
\label{equationFraction}
\end{equation}

By relying on spectral types to indicate the presence or absence of metals (Z) or hydrogen, we can identify four distinct observable states for He-atm~WDs: He-only, He+H, He+Z, and He+H+Z. If a population of $N$ stars contains $n$ in each state, and the fractional area of each cell in Fig.~\ref{fig:boxes} is represented by its label~(A--J), then the proportions of stars observed in each state are:

\begin{equation}
\begin{aligned}
n_{\textrm{He}}/N&=\textrm{A}+\textrm{F}\\
n_{\textrm{He+H}}/N&=\textrm{B}+\textrm{G}+\textrm{J}\\
n_{\textrm{He+Z}}/N&=\textrm{C}\\
n_{\textrm{He+H+Z}}/N&=\textrm{D}+\textrm{E}
\end{aligned}
\label{equationCounts}
\end{equation}

The parameters that determine the locations of the cell boundaries in Fig.~\ref{fig:boxes} are defined as follows: $p$ is the fraction of He-atm~WDs whose primordial hydrogen (if any) is currently detectable; $a$ is the fraction of He-atm~WDs that have ever detectably accreted material from their planetary systems (`accretors'); $d$ is the fraction of accretors with currently detectable metals; and $h$ is the fraction of accretors lacking in primordial hydrogen that have accumulated a detectable quantity of planetary hydrogen. The areas of cells A--J are fully determined by $a$, $d$, $h$ and $p$, so if we can find values for those parameters that reproduce the observed spectral type counts in equation~\ref{equationCounts}, then we can solve equation~\ref{equationFraction} to find $f$, the quantity of interest. We therefore use the cell areas to rewrite equations~\ref{equationFraction} and~\ref{equationCounts} in terms of the model parameters:

\begin{equation}
\begin{aligned}
\rm{A}&=(1-p)(1-a)\\
\rm{B}&=p(1-a)\\
\rm{C}&=(1-p)(1-h)ad\\
\rm{D}&=(1-p)had\\
\rm{E}&=pad\\
\rm{F}&=(1-p)(1-h)a(1-d)\\
\rm{G}&=(1-p)ha(1-d)\\
\rm{J}&=pa(1-d)\\
\end{aligned}
\label{equationCellAreas}
\end{equation}

\begin{equation}
f=\frac{(1-p)ha}{p+(1-p)ha}
\label{equationFractionRewritten}
\end{equation}

\begin{equation}
\begin{aligned}
n_{\textrm{He}}/N&=(1-p)((1-a)+a(1-d)(1-h))\\
n_{\textrm{He+H}}/N&=(1-a)p+a(1-d)(p+(1-p)h)\\
n_{\textrm{He+Z}}/N&=ad(1-p)(1-h)\\
n_{\textrm{He+H+Z}}/N&=ad((1-p)h+p)
\end{aligned}
\label{equationCountsRewritten}
\end{equation}

When all four parameters are free, multiple solutions for $f$ are possible. This can be seen from Fig.~\ref{fig:boxes}: while the data define the total areas of cells containing each type of white dwarf, their boundaries can still move (i.e. there are degeneracies between D and E, and between B, G and J). For example, to maintain the total area of cells A and F (pure He-atm~WDs), an increase in $p$ must be balanced by a decrease in $a$ and $d$. In the limiting case of $d=1$ (where all stars that have \textit{ever} accreted planetary material currently display metals in their spectra) the fraction of He-atm~WDs with detectable primordial hydrogen ($p$) is maximised, and thus $f$ is minimised.

\section{Sample selection}
\label{sec:sample}

We use spectral types to identify He-atm~WDs in catalogues of spectroscopically-confirmed white dwarfs based on DESI DR1 \citep{DESI2016science,DESI2025DR1,Swan2026} (44\,409 stars) and SDSS DR16 \citep{GentileFusillo2021,Ahumada2020} (29\,254 stars). The DESI survey is targeting all \textit{Gaia} white dwarfs within its footprint but outside the Galactic plane, down to a limit of $G\le20$\,mag \citep{Cooper2023}. We use only the primary sample (38\,904 stars) from the DESI DR1 white dwarf catalogue, which has a well-defined and largely unbiased selection function. The SDSS catalogue, by contrast, is drawn from numerous sub-surveys, with some specifically targeting white dwarfs (e.g. as faint high-proper-motion objects), and others observing them incidentally (e.g. in quasar programs). The SDSS white dwarf catalogue therefore has a complex selection function, and is thus more biased than the magnitude-limited DESI DR1 catalogue \edit{\citep[see fig.~12,][]{Swan2026}}.

Both SDSS and DESI may take multiple spectra of a given target. Spectral classifications are reported on a per-star basis in the DESI catalogue, but on a per-spectrum basis in the SDSS catalogue. Where a SDSS star has been assigned multiple spectral types, we use the classification from the spectrum with the highest signal-to-noise ratio (S/N), since in almost all such cases that is where the most features from metals or hydrogen are seen.

We consider only stars that are classified as DBs (those dominated by \ion{He}{i} transitions) and their subclasses, as a star showing features from both hydrogen and helium is likely to have a helium-dominated atmosphere, since hydrogen has a much higher opacity than helium. Exclusively relying on DB types avoids potential contamination from cool hydrogen-dominated stars, which can appear similar to He-atm~WDs with trace hydrogen below around 10\,000\,K, where He\,\textsc{i} lines disappear. However, it also potentially excludes He-atm WDs with sufficient hydrogen abundances that their spectra are dominated by Balmer lines (spectral type DA), but such instances are rare~--~less than 1 per cent of He-atm~WDs within 40\,pc \citep{OBrien2024_40pc}~--~and thus we do not expect any meaningful impact on our statistics from this choice.

Inevitably, some He-atm~WDs will have been misclassified, especially those with low-S/N spectra. We aim to mitigate such biases in our samples by enforcing minimum S/N requirements, and we also limit our selection to a colour range where all spectral features of interest are likely to have been identifiable. We now set out the S/N and colour criteria used to make our final selections. In Section~\ref{subsec:validation} we confirm that our results are not sensitive to the exact choices of these cuts.

Secondary spectral features from metals and hydrogen may be difficult to discern in low-S/N spectra, potentially inflating the value of $n_\mathrm{He}$. In each sample we identify the lowest-S/N He+H+Z-atm~WD, and exclude all stars with S/N below that threshold (S/N\,$< 3.9$ for DESI; S/N\,$< 13$ for SDSS). Note that we rely on the S/N values reported in each catalogue, which differ in their calculation methods, and thus are not comparable between the two surveys.

Separately, we apply a colour cut to exclude hot white dwarfs, where it becomes difficult to detect hydrogen or metals \citep{GentileFusillo2017}. We select objects with \textit{Gaia} colours $G_{\rm{BP}}-G_{\rm{RP}}>- 0.25$, corresponding to $T_{\textrm{eff}}\lesssim19\,000$\,K. Radiative levitation is negligible at such temperatures, so this cut has the additional benefit of unequivocally linking the presence of metals to accretion of planetary material.

Applying both the S/N and colour cuts to the catalogue He-atm~WDs yields 2407 and 1065 stars in the DESI and SDSS samples, respectively (Table~\ref{tab:data}). Of the 574 stars common to both, 95 are classified differently between the two catalogues, reflecting S/N differences: the stars where more features are detected in DESI have a higher average S/N in DESI and a lower average S/N in SDSS than the rest. While the two catalogues sample broadly the same region of the Hertzsprung--Russell (H--R) diagram, their underlying selection functions and S/N profiles differ sufficiently that we analyse the two samples independently. We use DESI as the primary sample as it is less biased. Thus, the SDSS sample serves as a comparison that illustrates how catalogue properties might bias the outcome.

\begin{table}
\centering
\caption{Spectral type counts used in our analysis.}
\label{tab:data}
\begin{tabular}{lll}
\hline
& DESI & SDSS\\
\hline
$n_\mathrm{He}$     & 1841 (76\,\%) & 765 (72\,\%) \\
$n_\mathrm{He+H}$   & 327  (14\,\%) & 174 (16\,\%) \\
$n_\mathrm{He+Z}$   & 102  (4\,\%)  & 58 (5\,\%)   \\
$n_\mathrm{He+H+Z}$ & 137  (6\,\%)  & 68 (6\,\%)   \\[5pt]
Total               & 2407          & 1065         \\
\hline
\end{tabular}
\end{table}

\section{Results and Implications}
\label{sec:results}

\subsection{A lower limit on the contribution of planetary hydrogen}
\label{subsec:min}

\begin{figure}
\centering
\includegraphics[width=0.85\columnwidth]{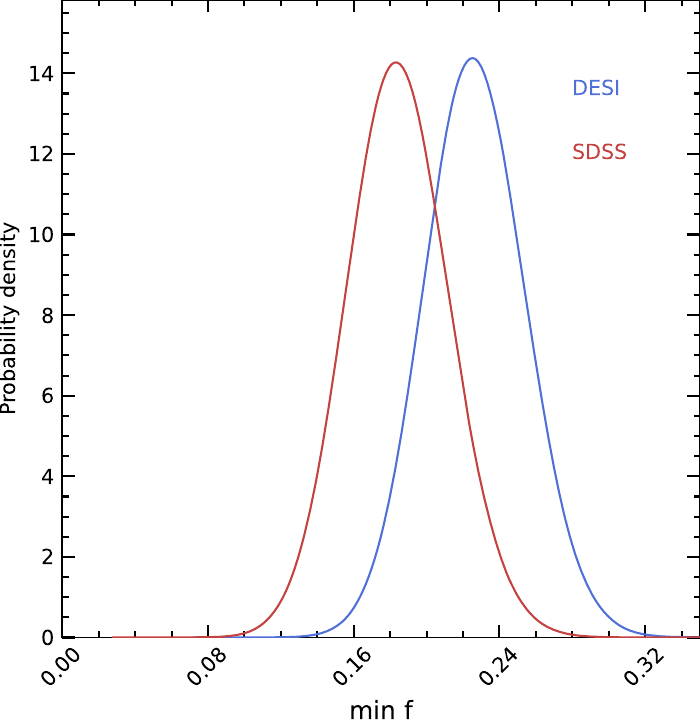}
\caption{The \textit{minimum} fraction $f$ of He+H~WDs that have accreted planetary hydrogen. Probability density distributions for $f$ based on each dataset are generated by bootstrapping.}
\label{fig:minf}
\end{figure}

\begin{figure}
\centering
\includegraphics[width=\hsize]{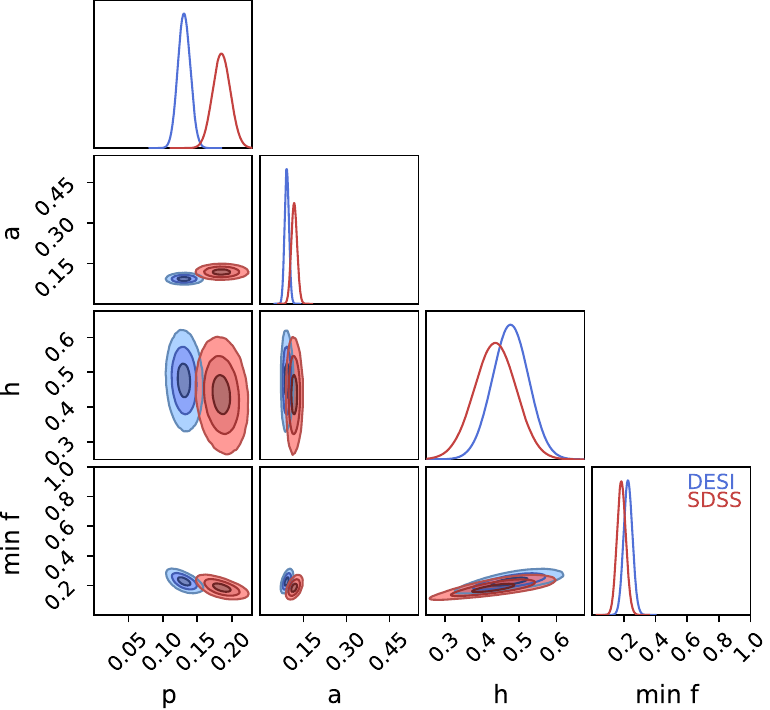}
\caption{Distributions of model parameters ($p$, $a$, $h$) and the \textit{minimum} fraction ($f$) of He+H-atm~WDs that have accreted planetary hydrogen from the bootstrapping analysis, where $p$ is the fraction of He-atm~WDs whose primordial hydrogen is detectable, $a$ is the fraction of He-atm~WDs that have ever accreted planetary material ('accretors'), and $h$ is the fraction of accretors lacking in primordial hydrogen that have detectable planetary hydrogen, in the conservative limit where the fraction of accretors with detectable metals is set to $d=1$.  Contours represent the 1-, 2- and 3-$\sigma$ regions of the distributions. NB this figure does \textit{not} represent a formal posterior probability distribution, but rather shows the range of solutions that are compatible with the bootstrapped data. Fig.~\ref{fig:minf} is a zoom into the lower-right panel of this plot.}
\label{fig:cornerLimits}
\end{figure}

We evaluate equation~\ref{equationFraction} to obtain the fraction of He+H-atm white dwarfs whose hydrogen includes a planetary contribution, using bootstrapping to estimate uncertainties. As stated above, to obtain a unique solution from this equation, one of the parameters must be held constant. We fix $d=1$ in order to minimise $f$, which is the most conservative scenario. This yields lower limits of $f\ge 0.22\pm0.02$ and $f\ge 0.18\pm0.03$ for the DESI and SDSS samples, respectively (Fig.~\ref{fig:minf}). The distributions for the remaining free parameters ($a$, $h$, $p$) are shown in Fig.~\ref{fig:cornerLimits} and summarised in Table~\ref{tab:parameters}.

\begin{table*}
\centering
\caption{Parameters of the model shown in Fig.~\ref{fig:boxes} in the conservative limit where all white dwarfs that have \textit{ever} accreted currently show detectable metals ($d=1$), and the corresponding inferred \textit{minimum} fraction $f$ of He+H-atm~WDs that have accreted planetary hydrogen.}
\label{tab:parameters}
\begin{tabular}{llll}
\hline
&& DESI & SDSS\\
\hline
Fraction of He-atm WDs that have ever detectably accreted planetary material & $a$          & $0.099\pm0.006$ & $0.12\pm0.01$ \\
Fraction of accretors with currently detectable metals                       & $d$ (assumed)& 1             & 1               \\
Fraction of accretors whose detectable hydrogen is planetary rather than primordial & $h$          & $0.50\pm0.04$ & $0.44\pm0.06$   \\
Fraction of He-atm WDs with detectable primordial hydrogen                   & $p$          & $0.151\pm0.008$ & $0.19\pm0.01$ \\[5pt]
Fraction of He+H-atm WDs whose hydrogen must include a planetary contribution       & $\min{f}$    & $0.22\pm0.02$ & $0.18\pm0.03$   \\
\hline
\end{tabular}
\end{table*}

\begin{figure}
\centering
\includegraphics[width=\hsize]{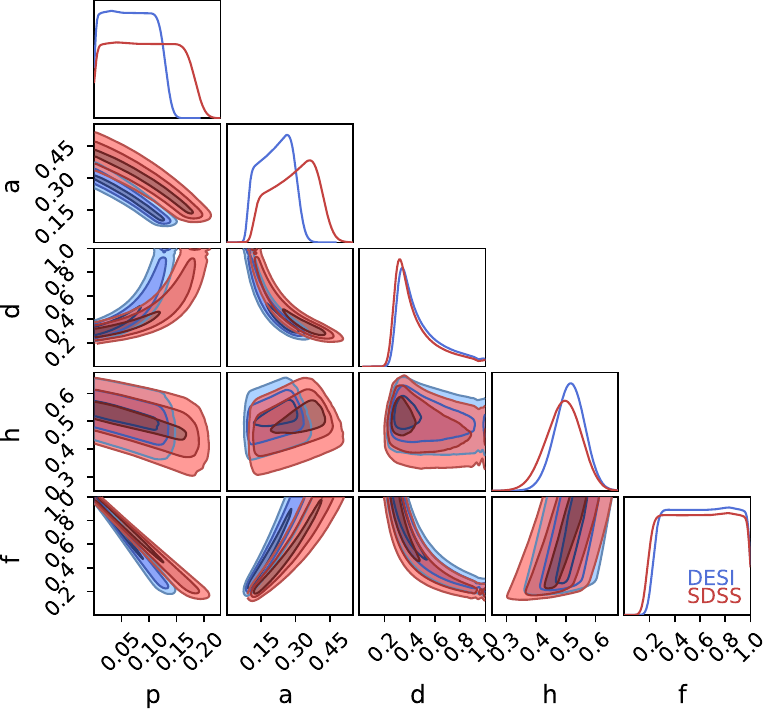}
\caption{Distributions of values compatible with the bootstrapped data when $d$ is free to vary, following Fig.~\ref{fig:cornerLimits}.}
\label{fig:cornerFull}
\end{figure}

Our lower limit on the fraction of He+H~WDs that have accreted planetary hydrogen ($f\ge 0.22\pm0.02$) is strikingly inconsistent with their being solely a product of stellar evolution ($f=0$). Instead, external delivery of hydrogen from planetary systems appears to be a necessary ingredient to explain the observed population of He+H-atm~WDs. Furthermore, this minimum represents the conservative and unlikely scenario where every member of the He+H-atm~WDs population that has ever experienced planetary accretion currently has metals in its atmosphere ($d=1$; no stars in cells F, G or J). However, only around 20--30 per cent of the He-atm white dwarfs currently show detectable metals \citep{Zuckerman2010,Manser2024HePlanets}. The true value of $f$ is therefore likely to be greater than the lower limits we find.

\subsection{Interpretation of model parameters}
\label{subsec:parameterInterpretation}

While our primary goal is to constrain $f$, the other parameters of our model may also provide useful insights. We therefore explore the full range of values for $p$, $a$, $d$, and $h$ that yield valid solutions. To do so we constrain $p$ to lie within each of 100 uniform slices of its possible range $0\leq p\leq n_{\textrm{He+H}}/(n_{\textrm{He}}+n_{\textrm{He+H}})$, and find a solution to equation~\ref{equationFractionRewritten} within each slice, using bootstrapped data. The resulting distributions are shown in Fig.~\ref{fig:cornerFull}.

The corner plot shows merely where the bootstrapped data permit solutions to our simple model, and it should not be over-interpreted. For example, solutions that entirely exclude primordial hydrogen ($p=0$) are in severe tension with current understanding of stellar evolution, and would be strongly disfavoured in a more sophisticated analysis. Furthermore, $p$ relates to primordial hydrogen \textit{currently detectable by visual inspection of spectra}, rather than a more fundamental physical quantity. It may vary along the cooling track, as primordial hydrogen wells up (increasing its detectability), convection zones develop (bringing former hydrogen-atmosphere white dwarfs into consideration as they become He+H-atm~WDs), and stars grow fainter (decreasing S/N, and thus the detectability of hydrogen). It is therefore specific to the samples we have selected rather than the whole He-atm~WD population. The forthcoming WEAVE survey \citep{Jin2024} will survey a similar number of white dwarfs to DESI, but at higher resolution, and thus will provide a useful comparison. Meanwhile, the 4MOST and SDSS-V surveys \citep{deJong2019,Kollmeier2025} will deliver comparable numbers of white dwarfs in the southern hemisphere.

While the $h$ parameter is well constrained, and suggests that nearly half of metal-accreting He-atm~WDs have acquired detectable quantities of planetary hydrogen, again we caution against over-interpretation. The $h$ parameter represents the fraction of systems that have accumulated a detectable quantity of hydrogen via planetary accretion, but says nothing about the nature of the material accreted during any given event, nor the timing of its arrival. Nevertheless, $h$ is largely insensitive to the other parameters, and to which survey the data are drawn from. Moreover, $h$ is significantly higher than the fraction of metal-enriched white dwarfs where accretion of ices is indicated \citep[only a handful of systems with oxygen detections show a 2-$\sigma$ excess above that expected from dry rock;][]{Brouwers2022waterContent}. This suggests that either most hydrogen is delivered in rare events involving large quantities of volatiles or hydrated minerals, or that the rocky material accreted by most stars is dry enough to avoid creating detectable oxygen excesses, but still contains enough trace hydrogen that it can accumulate to detectable levels over multiple accretion events.

As mentioned in Section~\ref{sec:model}, parameters $a$ and $d$ are coupled, so that only the fraction of accretors with detectable metals ($ad$) is observable, placing a lower limit on the frequency of remnant planetary systems. In our case, it is equivalent to the fraction of white dwarfs displaying metals features (i.e. a Z in their spectral type). From Table~\ref{tab:parameters} we find for the DESI and SDSS samples $ad = 0.10$ and $0.12$, respectively. These numbers are specific to our samples because metal detectability is temperature dependent. For example, the frequency measured from a different sample of cooler He-atm WDs in DESI EDR was $ad = 0.21$ \citep{Manser2024HePlanets}.

\subsection{Validation}
\label{subsec:validation}

\begin{figure*}
\centering
\includegraphics[width=\hsize]{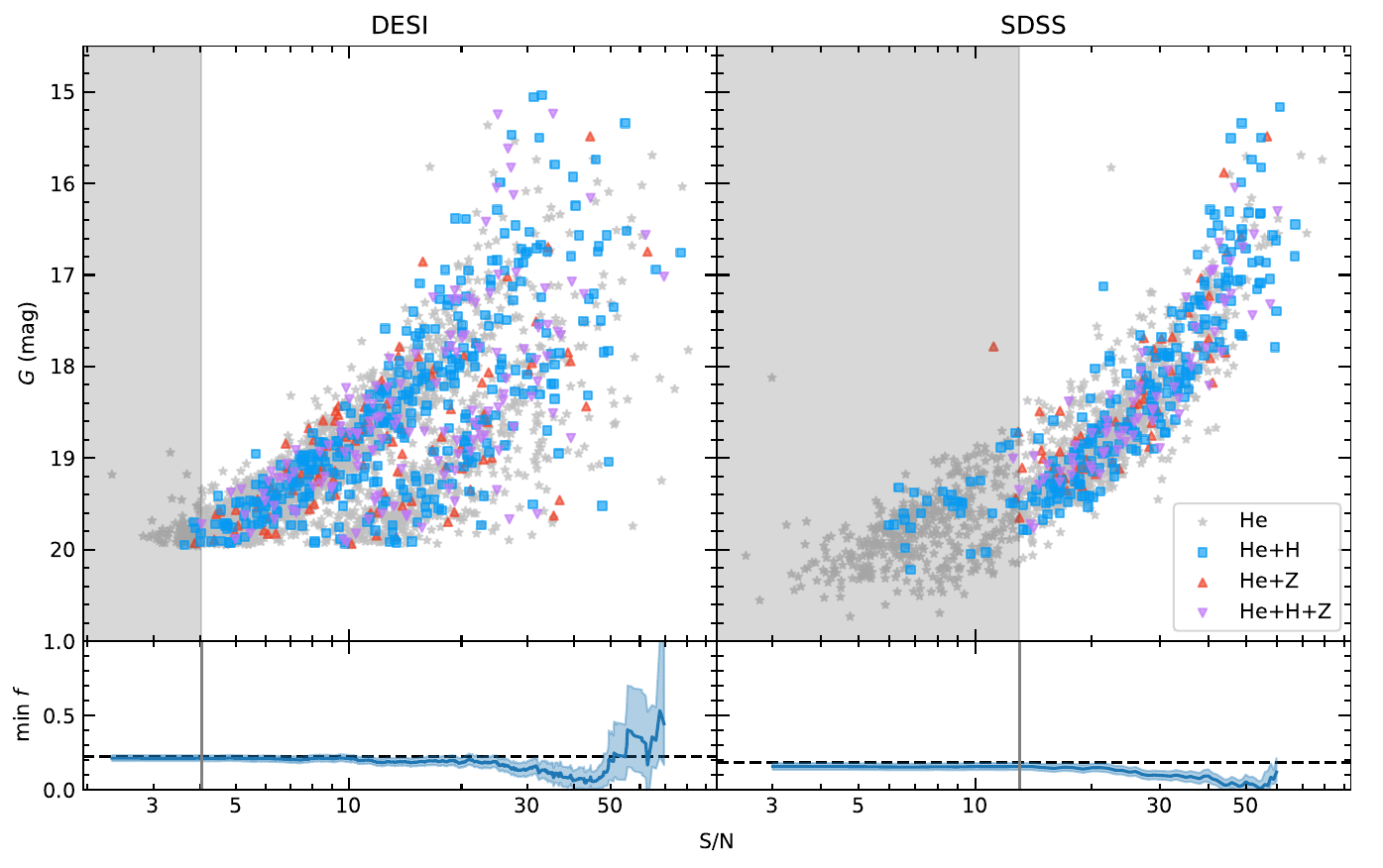}
\caption{Verification that results do not strongly depend on the choice of S/N cut. \textit{Top panels}:~S/N and \textit{Gaia G} magnitudes of candidate He-atm~WDs for inclusion in the samples, with the presence of metals (Z) or hydrogen indicated as in the legend. Shaded regions indicate our S/N cuts, corresponding to the lowest-S/N He+H+Z-atm~WD in each catalogue. The two tracks in the DESI plot reflect exposure time differences between dark- and bright-time observing programs. \textit{Bottom panels}:~Solid lines and shading show the \textit{minimum} value of $f$ and its 1-$\sigma$ uncertainty calculated after applying a cut on S/N\,$\ge x$ at each point on the $x$-axis, i.e. excluding all stars to the left of that point. The minimum value of $f$ calculated from our final samples are shown as dashed lines, with 1-$\sigma$ uncertainties as dotted lines.}
\label{fig:SN}
\end{figure*}

\begin{figure*}
\centering
\includegraphics[width=\hsize]{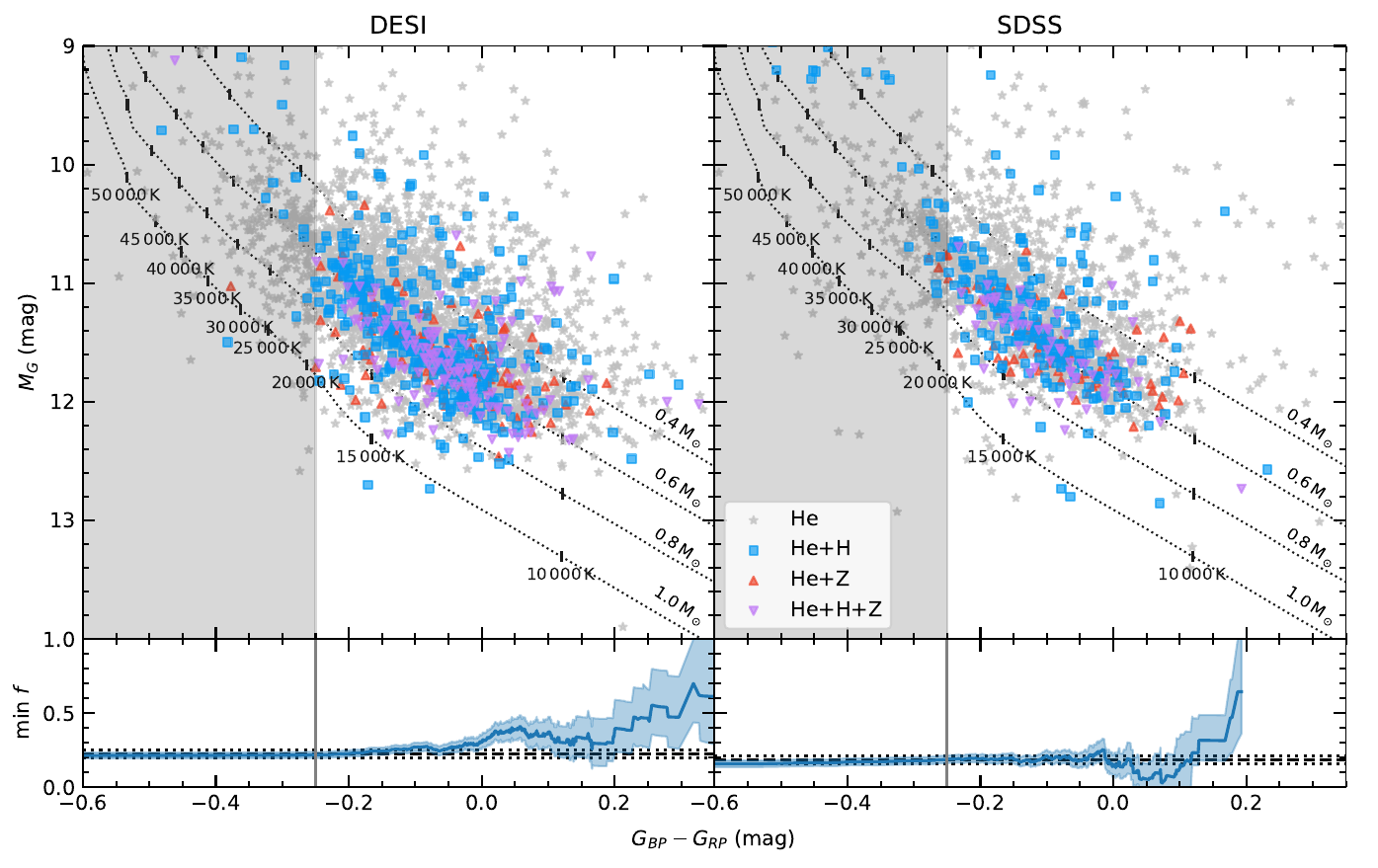}
\caption{Verification that results do not strongly depend on the choice of colour cut. \textit{Top panels}: Hertzsprung--Russell diagram showing candidate He-atm~WDs for inclusion in the samples. Symbols and shading have the same meaning as in Fig.~\ref{fig:SN}. Dotted lines show cooling tracks for pure He-atmosphere white dwarfs \protect \citep{Bedard2020} with masses in the range 0.4--1.0\,\Msun, with temperatures indicated by vertical marks. \textit{Bottom panels}: \textit{minimum} value of $f$ and its 1-$\sigma$ uncertainty calculated after applying a cut on $G_\mathrm{BP}-G_\mathrm{RP}\ge x$ at each point on the $x$-axis, following Fig.~\ref{fig:SN}.}
\label{fig:HRD}
\end{figure*}

Figs.~\ref{fig:SN} and \ref{fig:HRD} show the distribution of our samples in brightness, S/N and on the H--R~diagram, and quantify the effect on our results of varying the selection criteria specified in Section~\ref{sec:sample}.

The detectability of metals and hydrogen diminishes at low S/N, especially in the SDSS sample where many of the low-S/N spectra have been classified as He-atm~WDs, with only a handful of He+H-atm~WDs and no metal-enriched WDs at all at S/N\,$\leq10$. The bottom panels of Fig.~\ref{fig:SN} show the effect of imposing an increasingly-stringent minimum S/N cut on our calculations of the minimum value of $f$. For the most part, the value of $f$ is insensitive to the S/N threshold employed and remains consistent with the results from our final selections, deviating only toward the sparsely-populated high-S/N end. We are therefore confident that our choice of S/N cut has minimal impact on our result.

Candidate stars are well distributed across the colour--magnitude diagrams shown in Fig.~\ref{fig:HRD}, albeit with significant scatter. The bottom panels illustrate that $f$ is largely insensitive to the exact choice of our colour cut.

We caution against over-interpretation of the validation plots shown in Figs.~\ref{fig:SN} and \ref{fig:HRD}, specifically the potential variation of $\min{f}$ at the high-S/N and red ends of the bottom panels, where small-number statistics limit precision. \edit{We chose the cuts described in Section~\ref{sec:sample} to be maximally inclusive of detectable hydrogen and metals, and Figs.~\ref{fig:SN} and \ref{fig:HRD} are intended to demonstrate that our result is robust against small variations in those cuts. However, those figures also show that more substantial variations can impact the result, which we now discuss.}

\begin{figure}
\centering
\includegraphics[width=\hsize]{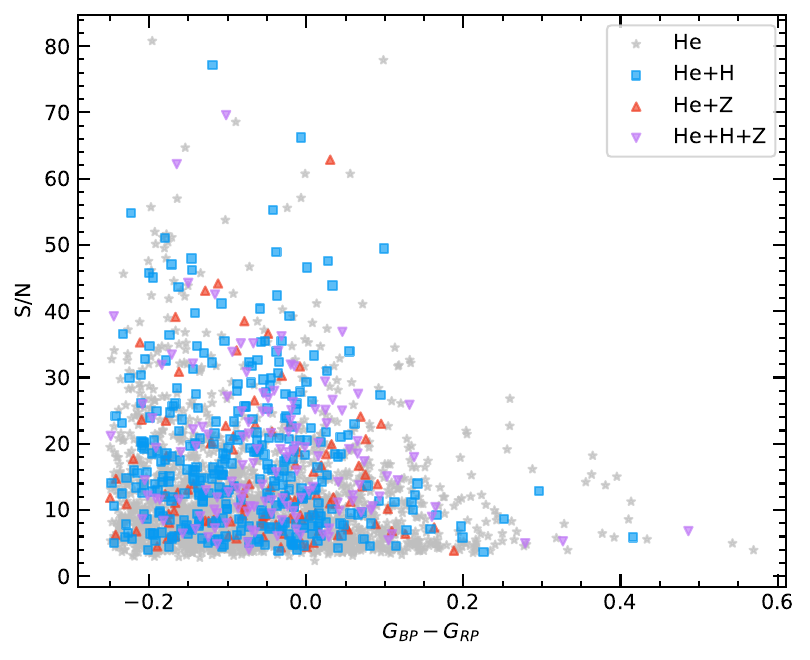}
\caption{\edit{The relationship between colour and S/N in the DESI sample, showing that \editmore{a more} stringent S/N cut than we made would bias the sample against redder (cooler, older) objects, changing its character.}}
\label{fig:SNBPRP}
\end{figure}

\begin{figure}
\centering
\includegraphics[width=\hsize]{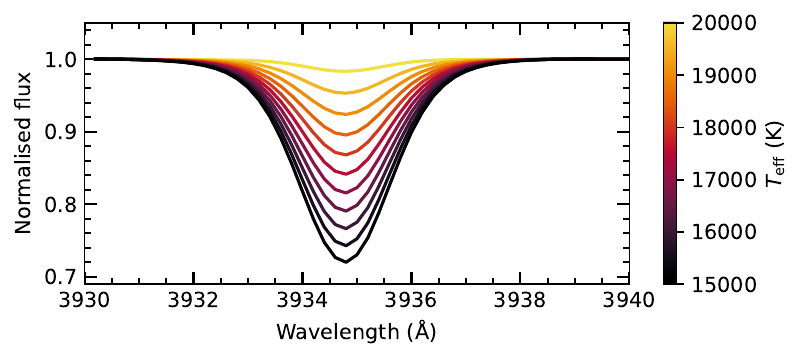}
\caption{\edit{The \ion{Ca}{ii} K line from simulated white dwarf spectra with the resolution and wavelength spacing of DESI across a range of effective temperatures. The other stellar parameters are fixed to $\log{g}=8.0$, $\log{\textrm{H}/\textrm{He}}=-5.0$ and $\log{\textrm{Ca}/\textrm{He}}=-8.0$, representing a moderate level of metal-enrichment.}}
\label{fig:CaTeff}
\end{figure}

\edit{While it might be tempting to make a more stringent S/N cut than we have used, to improve detectability of hydrogen and metals, that would alter the character of the sample.} Younger, hotter white dwarfs are bluer and more intrinsically luminous than older, cooler white dwarfs. In a magnitude-limited and uniformly-observed survey like DESI, hotter stars will be over-represented in a high-S/N sample, and under-represented in a red sample \edit{(Fig.~\ref{fig:SNBPRP}). For example, in the DESI sample, the median colour of the bottom five per cent by S/N is 0.1\,mag redder than that of the top five per cent, and Fig.~\ref{fig:HRD} suggests that $f$ may increase for a redder sample. Also, metals become harder to detect in hotter objects: the Ca\,K line weakens above 15\,000\,K as the ionisation state depopulates (Fig.~\ref{fig:CaTeff}), so a bluer sample will have a lower fraction of metal detections. Detectability of Balmer lines also varies with temperature, but to a lesser extent than Ca\,K in the range covered by our sample.}

\edit{Making a redder cut than we have used, to select only stars with strong \ion{Ca}{K} lines, would also change the character of the sample.} The very nature of white dwarf spectral evolution is that the fraction of He-atm~WDs with detectable hydrogen varies with cooling age\edit{, for which colour is a good proxy}. It is therefore not surprising to see some variation in $\min{f}$ as the sample is increasingly restricted to high-S/N or red objects, but an investigation of the cooling-age dependence of $f$ is well beyond the scope of this study.

\edit{White dwarfs in DESI DR1 have been analysed independently by multiple teams, so we assess the impact of classification methodology by repeating our analysis on a similar but independent catalogue of spectral types \citep{Kilic2026hot,Kilic2026cool}. We apply the method described in Section~\ref{sec:sample} to select a sample of 2360 objects from that catalogue. Evaluating equation~\ref{equationFraction} and fixing $d=1$ as described above, we find $f\ge0.16\pm0.02$. That limit is lower than ours at $2\upsigma$ significance, and may indicate that the subjectivity of spectral classification adds a systematic uncertainty of a few per cent to our result. However, the S/N profiles of the two DESI samples are different enough that they are not directly comparable: the DESI Collaboration team coadded all exposures for each object before classifying spectra \citep{Swan2026}, while the independent team inspected only one exposure per star, and 35 per cent of the stars in that sample have multiple exposures. Nevertheless, whichever catalogue of DESI spectral types is analysed, the result is still qualitatively the same: planetary hydrogen contributes significantly to white dwarf spectral evolution.}

\subsection{Assumptions and limitations}
\label{subsec:assumptions}

Our analysis makes three assumptions, which we justify below: (1)~delivery of planetary material has no dependence on the composition of the stellar atmosphere; (2)~spectral types are an unbiased reflection of atmospheric compositions; and (3)~stars evolve only along the paths shown in Fig.~\ref{fig:boxes}.

The rate of delivery of planetary material to a given type of white dwarf depends on three main factors: the frequency of planet formation around their progenitor stars, survival of those planetary systems through stellar evolution, and subsequent scattering of objects towards the white dwarf. The formation of He-atm~WDs depends sensitively on the timing of the final helium flash on the asymptotic giant branch \citep{Althaus2010}, rather than on bulk properties such as stellar mass. We therefore expect no correlation with the frequency of planet formation. While planets can be destroyed during giant branch evolution \citep{Villaver2007}, objects on orbits beyond several AU should remain far enough from the star that we expect their survival to have no meaningful dependence on the timing of helium flashes. Finally, the scattering of objects towards the white dwarf is a complex and unsolved dynamical problem, but if planetary system architectures are independent of atmospheric composition, then scattering rates should also be so. Metals are observed in 25--50 per cent of hydrogen-atmosphere white dwarfs \citep{Zuckerman2003,Koester2014,Wilson2019}, 30 per cent of warm He-atm~WDs \citep{Zuckerman2010}, and 21 per cent of cool He-atm~WDs \citep{Manser2024HePlanets}. The broad agreement between these statistics suggests that delivery rates do not dramatically differ within the narrower sub-classes of He-atm~WDs.

Metals and hydrogen can both contribute free electrons to a white dwarf atmosphere, potentially modifying photospheric opacities and absorption line strengths, and thus the presence of metals might inhibit or enhance the detectability of hydrogen, and vice versa. We therefore inspected model spectra across a range of abundances and temperatures to confirm that such effects are minimal, and of no concern in this context. \edit{The data support this: the fractions of stars with detected hydrogen and metals both increase with S/N (Table~\ref{tab:SN}). This contrasts with previous work on a smaller sample from SDSS, where only hydrogen was found to be detected more often at higher S/N, while the fraction of metal-enriched stars remained stable, albeit subject to uncertainty from small-number statistics \citep{Koester2015}.}

Detectability of metals and hydrogen could also be influenced by (i)~selection effects, (ii)~S/N ratios, and (iii)~spectral resolution. However, the two samples yield similar results, despite being drawn from surveys with different selection functions, S/N profiles, and resolutions. This suggests that systematic errors in the spectral classifications are not so strong as to invalidate our result. Moreover, the DESI~DR1 white dwarf catalogue is substantially less biased than its SDSS counterpart \citep{Swan2026}, reducing the risk of systematic effects.

\begin{table}
\centering
\caption{\edit{S/N dependence of the detectability of hydrogen and metals. The percentage of stars with an A or a Z in their spectral types (signifying features from hydrogen or metals, respectively) is calculated for several quantiles of the S/N distribution for each sample. Quantiles are used instead of absolute S/N values to enable comparison, as the measurements are survey-specific. Both hydrogen and metals are detected more often at higher S/N in each survey.}}
\label{tab:SN}
\begin{tabular}{lllll}
\hline
{} & \multicolumn{2}{c}{DBA(Z) \%} & \multicolumn{2}{c}{DB(A)Z \%} \\
S/N quantile & DESI & SDSS & DESI & SDSS \\
\hline
0.00 &       19 &   23 &       10 &   12 \\
0.50 &       27 &   31 &       13 &   14 \\
0.75 &       34 &   43 &       16 &   15 \\
0.85 &       38 &   49 &       19 &   18 \\
0.90 &       39 &   55 &       16 &   16 \\
\hline
\end{tabular}
\end{table}

Under our model white dwarfs cannot evolve leftwards (Fig.~\ref{fig:boxes}): based on the current model profiles of primordial hydrogen within helium atmospheres in white dwarfs \citep{Bedard2024}, the hydrogen present in such an atmosphere will always remain detectable, so a primordial He+H-atm~WD cannot evolve into a pure He-atm~WD. Likewise, a He-atm~WD within the temperature range of our samples would have shown detectable hydrogen of primordial origin by now, if present, and thus it cannot move from cell A to B.

\section{Conclusions}
\label{sec:conclusions}

Our empirical results show that a significant fraction of He+H-atm~WDs have accreted hydrogen from their planetary systems. In turn, this confirms that objects with significant hydrogen content often survive the post-main sequence evolution of their host stars \citep{Malamud2016waterI}. Meanwhile, state-of-the-art envelope models predict that white dwarfs born with a small quantity of primordial hydrogen will inevitably experience a He+H-atm phase \citep{Bedard2023}. These two sources of hydrogen are not mutually exclusive, so the challenge now is to disentangle their respective time-dependent contributions to white dwarf spectral evolution.

We have only used spectral types, but abundances of atmospheric hydrogen and metals, as well as cooling ages, can be inferred from white dwarf spectra. Modelling the evolution of primordial hydrogen in the envelopes provides time-dependent constraints on photospheric abundances attributable solely to mixing processes, and on the amount of hydrogen that a He+H-atm~WD can accrete without becoming hydrogen-dominated. Such constraints would facilitate distinguishing between scenarios such as capture of material from giant planets photo-evaporated by hot, young white dwarfs ($\sim$Myr), or stochastic accretion of ices throughout the whole post-main-sequence phase ($\sim$Gyr). Ultimately, interpreting hydrogen accretion histories in the context of photospheric metals and dynamical models could identify the source reservoirs of hydrogen-rich objects that are scattered towards white dwarfs. He+H-atm~WDs therefore have a pivotal role to play in mapping the locations of water-bearing planetesimals around white dwarfs and their progenitors across Galactic history, as well as in our understanding of the late stages of stellar evolution.

\section*{Acknowledgements}

\editmore{We thank the anonymous referee for a thorough report that helped to improve the paper.} We thank Antoine B\'{e}dard for useful discussions. This material is based upon work supported by the U.S. Department of Energy (DOE), Office of Science, Office of High-Energy Physics, under Contract No. DE–AC02–05CH11231, and by the National Energy Research Scientific Computing Center, a DOE Office of Science User Facility under the same contract. Additional support for DESI was provided by the U.S. National Science Foundation (NSF), Division of Astronomical Sciences under Contract No. AST-0950945 to the NSF’s National Optical-Infrared Astronomy Research Laboratory; the Science and Technology Facilities Council of the United Kingdom; the Gordon and Betty Moore Foundation; the Heising-Simons Foundation; the French Alternative Energies and Atomic Energy Commission (CEA); the Secretariat of Science, Humanities, Technology and Innovation (SECIHTI) of Mexico; the Ministry of Science, Innovation and Universities of Spain (MICIU/AEI/10.13039/501100011033), and by the \href{https://www.desi.lbl.gov/collaborating-institutions}{DESI Member Institutions}. Any opinions, findings, and conclusions or recommendations expressed in this material are those of the author(s) and do not necessarily reflect the views of the US National Science Foundation, the US Department of Energy, or any of the listed funding agencies. The authors are honored to be permitted to conduct scientific research on I'oligam Du'ag (Kitt Peak), a mountain with particular significance to the Tohono O’odham Nation. This project has received funding from the European Research Council (ERC) under the European Union’s Horizon 2020 research and innovation programme (Grant agreement No. 101020057). The authors acknowledge financial support from Imperial College London through an Imperial College Research Fellowship grant awarded to CJM. Corner plots were created using \href{https://samreay.github.io/ChainConsumer/}{ChainConsumer} \citep{Hinton2016}.

\textit{Author contributions}: CJM initiated the study. PI, AS, CJM and BG handled sample selection. PI, AS, CJM, BG and JN worked to develop the methods, with JN devising the geometric model on which our analysis is based. PI and AS wrote the article, contributing equally as joint first authors. AS and CJM produced the figures. The other authors have made significant contributions to the DESI Collaboration and the white dwarf survey underpinning this work.

\section*{Data Availability}

All data are available in the public catalogues cited. Data points shown in the figures are available in machine-readable format at \href{https://doi.org/10.5281/zenodo.14011623}{doi:10.5281/zenodo.14011623}.



\bibliographystyle{mnras}
\bibliography{DBAplanetaryH} 








\clearpage
\onecolumn
\section*{Affiliations}
$^{1}$Department of Physics, University of Warwick, Gibbet Hill Road, Coventry CV4 7AL, UK\\
$^{2}$Astrophysics Group, Department of Physics, Imperial College London, Prince Consort Rd, London SW7 2AZ, UK\\
$^{3}$NSF NOIRLab, 950 N. Cherry Ave., Tucson, AZ 85719, USA\\
$^{4}$Lawrence Berkeley National Laboratory, 1 Cyclotron Road, Berkeley, CA 94720, USA\\
$^{5}$Department of Physics, Boston University, 590 Commonwealth Avenue, Boston, MA 02215 USA\\
$^{6}$Department of Astronomy, University of Michigan, Ann Arbor, MI 48109, USA\\
$^{7}$Steward Observatory, University of Arizona, 933 N, Cherry Ave, Tucson, AZ 85721, USA\\
$^{8}$Dipartimento di Fisica ``Aldo Pontremoli'', Universit\`a degli Studi di Milano, Via Celoria 16, I-20133 Milano, Italy\\
$^{9}$INAF-Osservatorio Astronomico di Brera, Via Brera 28, 20122 Milano, Italy\\
$^{10}$Department of Physics \& Astronomy, University College London, Gower Street, London WC1E 6BT, UK\\
$^{11}$Department of Physics and Astronomy, The University of Utah, 115 South 1400 East, Salt Lake City, UT 84112, USA\\
$^{12}$Instituto de F\'{\i}sica, Universidad Nacional Aut\'{o}noma de M\'{e}xico,  Circuito de la Investigaci\'{o}n Cient\'{\i}fica, Ciudad Universitaria, Cd. de M\'{e}xico  C.~P.~04510,  M\'{e}xico\\
$^{13}$Instituci\'{o} Catalana de Recerca i Estudis Avan\c{c}ats, Passeig de Llu\'{\i}s Companys, 23, 08010 Barcelona, Spain\\
$^{14}$Institut de F\'{i}sica dâ€™Altes Energies (IFAE), The Barcelona Institute of Science and Technology, Edifici Cn, Campus UAB, 08193, Bellaterra (Barcelona), Spain\\
$^{15}$Departamento de F\'isica, Universidad de los Andes, Cra. 1 No. 18A-10, Edificio Ip, CP 111711, Bogot\'a, Colombia\\
$^{16}$Observatorio Astron\'omico, Universidad de los Andes, Cra. 1 No. 18A-10, Edificio H, CP 111711 Bogot\'a, Colombia\\
$^{17}$Institut d'Estudis Espacials de Catalunya (IEEC), c/ Esteve Terradas 1, Edifici RDIT, Campus PMT-UPC, 08860 Castelldefels, Spain\\
$^{18}$Institute of Cosmology and Gravitation, University of Portsmouth, Dennis Sciama Building, Portsmouth PO1 3FX, UK\\
$^{19}$Institute of Space Sciences, ICE-CSIC, Campus UAB, Carrer de Can Magrans s/n, 08913 Bellaterra, Barcelona, Spain\\
$^{20}$University of Virginia, Department of Astronomy, Charlottesville, VA 22904, USA\\
$^{21}$Fermi National Accelerator Laboratory, PO Box 500, Batavia, IL 60510, USA\\
$^{22}$Center for Cosmology and AstroParticle Physics, The Ohio State University, 191 West Woodruff Avenue, Columbus, OH 43210, USA\\
$^{23}$Department of Physics, The Ohio State University, 191 West Woodruff Avenue, Columbus, OH 43210, USA\\
$^{24}$The Ohio State University, Columbus, 43210 OH, USA\\
$^{25}$Department of Physics, Southern Methodist University, 3215 Daniel Avenue, Dallas, TX 75275, USA\\
$^{26}$Institute for Astronomy, University of Edinburgh, Royal Observatory, Blackford Hill, Edinburgh EH9 3HJ, UK\\
$^{27}$Institute of Astronomy, University of Cambridge, Madingley Road, Cambridge CB3 0HA, UK\\
$^{28}$Sorbonne Universit\'{e}, CNRS/IN2P3, Laboratoire de Physique Nucl\'{e}aire et de Hautes Energies (LPNHE), FR-75005 Paris, France\\
$^{29}$Department of Astronomy \& Astrophysics, University of Toronto, Toronto, ON M5S 3H4, Canada\\
$^{30}$Department of Physics and Astronomy, Siena University, 515 Loudon Road, Loudonville, NY 12211, USA\\
$^{31}$Instituto de Astrof\'{i}sica de Andaluc\'{i}a (CSIC), Glorieta de la Astronom\'{i}a, s/n, E-18008 Granada, Spain\\
$^{32}$Departament de F\'isica, EEBE, Universitat Polit\`ecnica de Catalunya, c/Eduard Maristany 10, 08930 Barcelona, Spain\\
$^{33}$Institute for Computational Cosmology, Department of Physics, Durham University, South Road, Durham DH1 3LE, UK\\
$^{34}$Department of Physics and Astronomy, Sejong University, 209 Neungdong-ro, Gwangjin-gu, Seoul 05006, Republic of Korea\\
$^{35}$CIEMAT, Avenida Complutense 40, E-28040 Madrid, Spain\\
$^{36}$Department of Physics, University of Michigan, 450 Church Street, Ann Arbor, MI 48109, USA\\
$^{37}$University of Michigan, 500 S. State Street, Ann Arbor, MI 48109, USA\\
$^{38}$National Astronomical Observatories, Chinese Academy of Sciences, A20 Datun Road, Chaoyang District, Beijing, 100101, P.~R.~China\\

\bsp	
\label{lastpage}
\end{document}